\documentclass[aps,pra,superscriptaddress,twocolumn,floatfix]{revtex4-2}
\usepackage{amstext}
\usepackage{amsmath}
\usepackage{amsfonts}
\usepackage{bm}
\usepackage{graphicx}
\usepackage{graphics}
\usepackage{subfigure}
\usepackage{esint}
\usepackage{color}
\usepackage{physics}
\usepackage[normalem]{ulem}
\usepackage[colorlinks, linkcolor=blue, citecolor=blue,hyperindex,bookmarks=false,pdfstartview=FitH]{hyperref}
\usepackage{lineno}
\begin{document}

\title{Pulsed Generation of Continuous-Variable Cluster States in a Phononic Quantum Network}

\author{Anuvetha Govindarajan}
\affiliation{School of Natural Sciences, University of California, Merced, 5200 North Lake Road, Merced, California, 95343, USA}

\author{Mitch Mazzei}
\affiliation{School of Natural Sciences, University of California, Merced, 5200 North Lake Road, Merced, California, 95343, USA}

\author{Hailin Wang}
\affiliation{Department of Physics, University of Oregon, Eugene, Oregon 97403, USA}

\author{Lin Tian}
\email{ltian@ucmerced.edu}
\affiliation{School of Natural Sciences, University of California, Merced, 5200 North Lake Road, Merced, California, 95343, USA}

\begin{abstract}
Cluster states are multipartite entangled states that are maximally connected and resilient to decoherence, making them valuable resources for quantum information processing. Continuous-variable (CV) cluster states have been extensively investigated for such applications.  Here we present a pulsed protocol for generating CV cluster states in a phononic quantum network composed of phonon waveguides, mechanical resonators, and optical cavities. A key feature of this architecture is its modular design, where pairs of mechanical modes serve as building blocks with only local, tunable interactions between mechanical and cavity modes. We characterize the generated cluster states through the average nullifier of the CV modes. Our numerical results show that high-quality CV cluster states can be generated with moderate squeezing parameters, whereas strong squeezing and the resulting large phonon occupations can degrade the cluster states in the presence of finite dissipation. We further show that, under moderate squeezing and dissipation, the average nullifier can decrease with increasing system size $N$, demonstrating the scalability of the proposed scheme. As a direct application, we demonstrate that distant mechanical modes can be entangled through local measurements.
\end{abstract}
\maketitle

\section{Introduction \label{sec:intro}}
Quantum networks are central to distributed quantum computing, secure communication, and scalable quantum technologies~\cite{CiracPRL1997}. Among the diverse architectures under development, phononic quantum networks have emerged as a promising platform owing to their long coherence times and compatibility with disparate quantum systems, which enable coherent interfaces between fundamentally different platforms such as superconducting and optical devices~\cite{GustafssonPropagatingAtom, LemondePRL2018PhononWaveguides, KuzykPRX2018, MaityNatComm2020CoherentDiamond, XLiHWangNanoLett2024, DiamandiNatPhys2025, ArrazolaPRB2024,  RPeng2025arXiv2409_12938}. 
Acoustic resonators with quality factors reaching $10^{11}$~\cite{YWangLaserPhys2023} are particularly attractive as long-lived quantum memories and coherent transducers. Optomechanical systems, which exploit radiation-pressure interactions between cavity and mechanical modes, provide a powerful route for engineering nonclassical states of both phonons and photons~\cite{AspelmeyerRMP2014CavityOptomechanics,Meystre2013AnnPhysOptomechanics}. In particular, multimode optomechanical systems have been widely investigated for the generation of quantum many-body states~\cite{BhattacharyaPRA2008, MasselNatCommun2012, LudwigPRL2013, BarzanjehNatCommun2017, NielsenPNAS2017, RuesinkNatCommun2018, TianPRA2017, KharelPRAppl2022,YoussefiNature2022}, and optomechanical entanglement has been demonstrated across multiple regimes and experimental platforms~\cite{Ockeloen-KorppiNature2018, JiangPRA2020, ZippilliPRL2021, KotlerScience2021, ThomasNatPhys2021, HQianQST2023}.

A key class of quantum resources for such quantum networks is cluster states
~\cite{Briegel2001PersistentParticles}, which are multipartite entangled states that are maximally connected and resilient to loss. These states form the foundation of measurement-based quantum computation~\cite{Raussendorf2001AComputer, WaltherNature2005}. While discrete-variable (DV) cluster states have been extensively explored, continuous-variable (CV) cluster states provide a powerful alternative platform~\cite{WeedbrookRMP2012GaussianQI, AdessoOpenSystInfDyn2014}. 
CV cluster states can be generated using off-line squeezing in combination with linear quantum operations~\cite{ZhangPRA2006, Menicucci2006, vanLoockPRA2007, GuPRA2009, ZippilliVitaliPRA2020}. Large-scale implementations of CV cluster states have been experimentally demonstrated in multiple degrees of freedom in photonic systems, including the time domain with more than 10,000 modes~\cite{YokoyamaNatPhoton2013, YoshikawaAPLPhotonics2016, AsavanantScience2019,LarsenScience2019} and the frequency domain~\cite{ChenPRL2014ExperimentalComb, RoslundNatPhotonics2014, ZWang2024largescaleclusterquantummicrocombs}. Additional proposals extend to spatial~\cite{PooserPRA2014Continuous-variableComb}, temporal~\cite{DSuPRA2018}, and time-frequency encodings~\cite{AlexanderPRA2016, HumphreysPRL2014}, as well as hybrid DV-CV architectures~\cite{OmkarPRL2020, WuPRR2020QuantumPlatform}. More recently, CV cluster states have also been proposed in microwave systems~\cite{YazdiNJP2023,LinguaPRL2025}. 
Meanwhile, generating such states with mechanical degrees of freedom, including vibrational modes in trapped ions~\cite{YLiYHLinScienceAdvances2025TrappedIon} and optomechanical systems, offers a promising route toward scalable and integrated quantum architectures. Approaches based on reservoir engineering in dissipative systems~\cite{HouhouPRA2015CVClusterstateOptomechanics, MoorePRA2017, YazdiVitaliQST2024} and entanglement swapping~\cite{OttavianiPRA2019MultipartiteStates} have been investigated. 

Here we present a pulsed protocol for generating large-scale CV cluster states in a modular optomechanical quantum network, where each mechanical resonator is coupled to a cavity mode. The architecture of this quantum network is inspired by the modular configuration of a phononic quantum network studied in Ref.~\cite{KuzykPRX2018}, in which acoustic waveguides isolate neighboring subsystems and enable effective local operations without inducing crosstalk between distant modes. Such a modular architecture provides a natural route toward scalable quantum engineering. The tunability of the system allows both two-mode squeezing and beam-splitter operations to be implemented through local radiation-pressure couplings between cavities and mechanical resonators. We derive the symplectic matrix associated with these operations and obtain the corresponding adjacency matrix that defines the graph of the generated cluster state. The quality of the cluster state is characterized by evaluating the average nullifier of the mechanical modes. We further analyze the effects of mechanical and cavity dissipation using the Heisenberg--Langevin equations. Our results show that high-quality CV cluster states can be generated with moderate squeezing parameters, whereas strong squeezing and the resulting large phonon occupations can degrade the cluster states in the presence of finite dissipation. We further show that, under moderate squeezing and dissipation, the average nullifier can decrease with increasing system size, demonstrating the scalability of the proposed protocol. As a direct application, we demonstrate that distant mechanical modes can be entangled through local measurements within the CV cluster state. Importantly, the scheme requires only $O(N)$ driving tones for a quantum network with $N$ mechanical resonators. These results establish hybrid phononic quantum networks as a promising platform for scalable continuous-variable quantum technologies.

Our work complements previous approaches for generating mechanical CV cluster states via reservoir engineering~\cite{HouhouPRA2015CVClusterstateOptomechanics, MoorePRA2017, YazdiVitaliQST2024}, where cluster states are generated dissipatively as steady or quasi-steady states. In contrast, the present pulsed protocol exploits the modular architecture of the phononic network, in which interactions are implemented through programmable pulsed operations. This architecture naturally suppresses crosstalk between different subsystems and provides a scalable route toward the generation of large cluster states. The pulsed approach also offers flexible dynamical control over the generated Gaussian states and their associated graph structures. The trade-off is that the present protocol operates in the strong-coupling regime, which can be experimentally more demanding than the weak-coupling regime typically employed in reservoir-engineering approaches.

The paper is organized as follows. In Sec.~\ref{sec:system}, we describe the phononic quantum network and the quantum operations induced by strong driving of the cavity modes. In Sec.~\ref{sec:protocol}, we introduce a three-step protocol for generating CV cluster states, derive the corresponding symplectic and adjacency matrices, and obtain the average nullifier as a measure of the quality of the generated states. In Sec.~\ref{sec:decoherence}, we analyze the effects of mechanical and cavity dissipation using the Heisenberg--Langevin equations and investigate the scalability of the protocol under finite dissipation. In Sec.~\ref{sec:entanglementtransfer}, we demonstrate that distant mechanical modes can be entangled through local quadrature measurements. Finally, conclusions are presented in Sec.~\ref{sec:conclusions}.

\section{System \label{sec:system}}
\begin{figure} 
\includegraphics[width=8cm, clip]{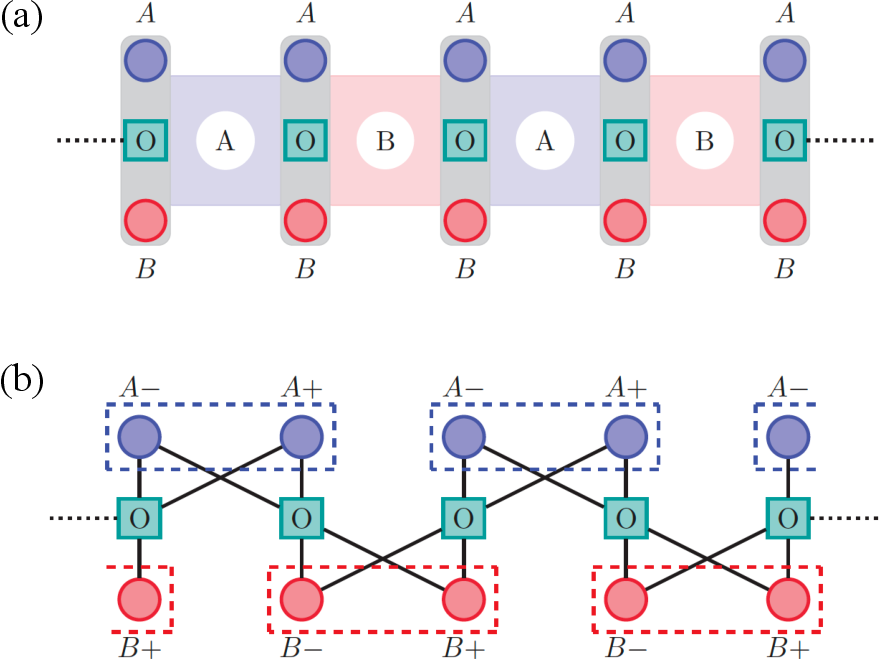}
\caption{(a) Schematic of the phononic quantum network. Each mechanical resonator (rectangular block) hosts two mechanical modes, $A$ (blue circle) and $B$ (red circle), coupled to a single cavity mode $O$. Neighboring resonators are connected by phononic waveguides A (blue block) or B (red block).  (b) Coupling structure of the quantum network. The mechanical normal modes $A_{\pm}$ (upper modes) and $B_{\pm}$ (lower modes) form isolated subsystems, indicated by the dashed boxes. Solid lines denote the couplings between the normal modes and the cavity modes.}
\label{fig1}
\end{figure}

The phononic quantum network consists of an array of mechanical resonators interconnected by phononic waveguides, similar to the configuration in \cite{KuzykPRX2018}. Each resonator hosts two mechanical modes with well-separated frequencies, $\omega_{A}$ (upper mode) and $\omega_{B}$ (lower mode), together with an embedded cavity, as illustrated in Fig.~\ref{fig1}(a). The mechanical modes couple to the cavity through radiation-pressure interactions, while neighboring resonators are connected by phononic waveguides that alternatively support phonon propagation near $\omega_{A}$ or $\omega_{B}$. Consequently, the discrete standing-wave mode of waveguide A (B) and the mechanical modes at frequency $\omega_A$ ($\omega_B$) in adjacent resonators form an isolated subsystem with three hybridized normal modes. From each subsystem, we select two normal modes, denoted by $A_{\pm}$ ($B_{\pm}$), with frequencies $\omega_{A_{\pm}}=\omega_A\pm \Lambda_A/2$ ($\omega_{B_{\pm}}=\omega_B\pm \Lambda_B/2$), to construct the CV cluster states. The frequency splittings $\Lambda_A$ and $\Lambda_B$ are determined by the coupling strengths between the resonator and waveguide modes as well as their detunings. Each selected normal mode is a superposition of local mechanical and waveguide modes and couples to the  cavity mode, as shown in Fig.~\ref{fig1}(b).

The Hamiltonian of this system takes the form $H_{\rm t} = H_{\rm m} + H_{\rm c} + H_{\rm r}$ ($\hbar\equiv1$), where
\begin{eqnarray}
    H_{\rm m} &=& \sum_{n={\rm odd}} \left( \omega_{A_{-}} b_{n}^\dag b_{n} + \omega_{B_{+}} b_{N+n}^\dag b_{N+n}\right) \nonumber \\
    && + \sum_{n={\rm even}} \left( \omega_{A_{+}} b_{n}^\dag b_{n} + \omega_{B_{-}} b_{N+n}^\dag b_{N+n}\right) 
    \label{eq:Hm}    
\end{eqnarray}
is the Hamiltonian of the mechanical normal modes, $H_{\rm c} = \sum_n \omega_{c}  a_n^\dagger a_n$ is the Hamiltonian of the cavity modes, $H_{\rm r} = \sum_{\langle n,m\rangle} g_{\alpha} a_n^\dagger a_n (b_{m}+b_{m}^\dag)$ describes the radiation-pressure interaction, and the summation $\langle n,m\rangle$ runs over all allowed couplings between the cavity and mechanical modes. 
Here, $b_n$ and $b_{N+n}$ ($b_n^\dagger$ and $b_{N+n}^\dagger$) are the annihilation (creation) operators of the upper and lower mechanical modes at site $n$, respectively, where $n\in[1,N]$ and $N$ is the number of resonators in the array. The operators $a_n$ ($a_n^\dagger$) are the annihilation (creation) operators of the cavity mode at site $n$, $\omega_c$ is the cavity frequency, and $g_\alpha$ ($\alpha=A,B$) denotes the single-photon optomechanical coupling strength. 
For a mechanical resonator at an odd (even) site $n$, the upper mechanical mode $b_n$ corresponds to an $A_{-}$ ($A_{+}$) mode, while the lower mechanical mode $b_{N+n}$ corresponds to a $B_{+}$ ($B_{-}$) mode. The cavity mode at an odd site $n$ couples to the mechanical modes $b_n$, $b_{n+1}$, $b_{N+n-1}$, and $b_{N+n}$, whereas the cavity mode at an even site couples to $b_{n-1}$, $b_n$, $b_{N+n}$, and $b_{N+n+1}$, as shown in Fig.~\ref{fig1}(b). 
For simplicity, we assume that the coupling strengths $g_\alpha$, as well as the cavity and mechanical frequencies, are site-independent. These parameters can, however, be engineered to be site dependent if desired.

When the energy separations between the mechanical normal modes are much larger than the cavity linewidth $\kappa$, i.e., $\Lambda_A$, $\Lambda_B$, and $|\Lambda_A-\Lambda_B| \gg \kappa$, strong driving fields can be applied to the cavities to linearize the radiation-pressure interactions and generate desired linear or bilinear couplings between the cavity and mechanical modes~\cite{TianReview2015}.
For example, with a driving frequency $\omega_d = \omega_c -\omega_{A_{-}}$ (the first red sideband of mode $A_{-}$), the linearized interaction at an odd site $n$ is $H_{\rm I} = g_I a_n^\dagger  b_{n} + h.c.$ with the coupling strength $g_I = g_{A} \langle a_n^\dagger a_n \rangle^{1/2}$. Therefore a beam-splitter operation between the cavity mode $a_n$ and the mechanical mode $b_{n}$ is generated. With a driving frequency $\omega_d = \omega_c + \omega_{A_{-}}$ (the first blue sideband), a parametric amplification operation with $H_{I} = g_I a_n^\dagger  b_{n}^\dagger + h.c.$ can be obtained. 
These linearized interactions enable coherent manipulation of the mechanical modes in the phononic quantum network and support a broad range of quantum protocols. The high degree of tunability of the optomechanical interactions facilitates the implementation of beam-splitter, single-mode squeezing, and two-mode squeezing operations, which are essential for generating entanglement and cluster states in CV systems. Moreover, because the phononic waveguides selectively couple mechanical modes into isolated subsystems, crosstalk between different building blocks is effectively suppressed. As a result, the protocol can be naturally scaled to large quantum networks connecting distant nodes.

\section{Protocol \label{sec:protocol}}
The phononic quantum network with $N$ mechanical resonators contains $2N$ mechanical normal modes. We denote the displacement quadrature of the $n$th mode by $Q_n=(b_n+b_n^\dagger)/\sqrt{2}$ and the momentum quadrature by $P_n=-i(b_n-b_n^\dagger)/\sqrt{2}$, where $n\in[1,2N]$. In the ideal case, a CV cluster state is generated by applying the unitary transformation $U=\prod_{m,n}\exp\!\left(\frac{i}{2}A_{mn}Q_mQ_n\right)$ to the zero momentum state, where $A_{mn}$ are the elements of the adjacency matrix $\mathbf{A}$~\cite{ZhangPRA2006,Menicucci2006}. To characterize the state, we define a set of nullifier operators $N_{n} = P_{n} -\sum_{m} A_{nm} Q_{m}$. The CV cluster state $\ket{\psi}$ associated with the adjacency matrix $\mathbf{A}$ satisfies $N_n\ket{\psi}=0$. The entanglement structure of the CV cluster state can be represented by a graph defined by $\mathbf{A}$~\cite{MenicucciPRA2011_1,ZhangPRA2008_1}, where the vertices correspond to the CV modes and the weighted edges encode their correlations.
In realistic systems, an ideal zero momentum state is not available. Instead, approximate CV cluster states can be generated by applying squeezing and linear optical operations to an initial Gaussian state. In the limit of infinite squeezing, the resulting Gaussian state approaches the ideal CV cluster state with 
${\rm cov}(N_n)_{\alpha\rightarrow\infty}=0$ for all $n$, where $\alpha$ is the squeezing parameter of the initial state.

In the following, we introduce a three-step pulsed protocol~\cite{GovindarajanThesis2025} for generating approximate CV cluster states in the phononic quantum network described in Sec.~\ref{sec:system}. The protocol exploits the intrinsic tunability of the network by combining pulsed beam-splitter and parametric-amplification operations between the mechanical and cavity modes. For an array of $N$ mechanical resonators, corresponding to $2N$ mechanical modes, the scheme requires $(4N-2)$ driving tones.

\subsection*{Step I}
In the first step, we generate momentum squeezing on all $2N$ mechanical modes. We begin by applying a beam-splitter interaction between the lower mechanical mode $b_{N+n}$ and the cavity mode $a_n$, governed by the Hamiltonian $H_{\rm I}=-ig_I(b_{N+n}^\dagger a_n-a_n^\dagger b_{N+n})$. After an interaction time $t_{10}=\pi/2g_I$, the two modes are swapped, yielding the cavity quadratures $Q_{an}(t_{10})=Q_{N+n}(0)$ and $P_{an}(t_{10})=P_{N+n}(0)$, and the mechanical quadratures $Q_{N+n}(t_{10})=-Q_{an}(0)$ and $P_{N+n}(t_{10})=-P_{an}(0)$.
Next, we apply a parametric-amplification interaction between the upper mechanical mode $b_n$ and the cavity mode $a_n$, described by the Hamiltonian $H_{\rm I}=-g_I(b_n^\dagger a_n^\dagger+a_n b_n)$. This operation generates two-mode squeezing between $b_n$ and $a_n$. At the time $t_{11}=t_{10}+\alpha/g_I$, we obtain
\begin{subequations}
    \begin{align}
        P_n(t_{11}) - Q_{an}(t_{11}) &= e^{-\alpha} \big[ P_n(0) - Q_{N+n}(0) \big], \\ 
        P_{an}(t_{11}) - Q_{n}(t_{11}) &= e^{\alpha} \big[ P_{N+n}(0) - Q_{n}(0) \big],
    \end{align}
\end{subequations}
where $\alpha$ is the squeezing parameter of the initial state.
Next, a beam-splitter interaction between $b_n$ and $a_n$, governed by $H_{\rm I}=g_I(b_n^\dagger a_n+a_n^\dagger b_n)$, is applied for a duration $\pi/4g_I$. At the time $t_{12}=t_{11}+\pi/4g_I$, the two-mode squeezing is converted into single-mode squeezing.
Finally, a second beam-splitter interaction between $b_{N+n}$ and $a_n$, governed by $H_{\rm I}=ig_I(b_{N+n}^\dagger a_n-a_n^\dagger b_{N+n})$, is applied for a duration $\pi/2g_I$. This operation transfers the single-mode squeezing from $a_n$ back to $b_{N+n}$. At the final time $\tau_1=t_{12}+\pi/2g_I$, the quadratures become
\begin{subequations}
    \begin{align}
        Q_{n}(\tau_1) &= \tfrac{e^\alpha}{\sqrt{2}} \big[ Q_{n}(0) + P_{N+n}(0) \big], \\
        Q_{N+n}(\tau_1) &= \tfrac{e^\alpha}{\sqrt{2}} \big[ Q_{N+n}(0) + P_{n}(0) \big], \\
        P_{n}(\tau_1) &= \tfrac{e^{-\alpha}}{\sqrt{2}} \big[ P_{n}(0) - Q_{N+n}(0) \big], \\ 
        P_{N+n}(\tau_1) &= \tfrac{e^{-\alpha}}{\sqrt{2}} \big[ P_{N+n}(0) - Q_{n}(0) \big].
    \end{align}
\end{subequations}
As a result, all $2N$ mechanical modes are prepared in momentum-squeezed states, with covariances ${\rm cov}(P_n)=e^{-2\alpha}$ and ${\rm cov}(Q_n)=e^{2\alpha}$. At the same time, the cavity modes are restored to their initial vacuum states and are completely disentangled from the mechanical modes.

\begin{figure}
    \centering
    \includegraphics[width=8cm, clip]{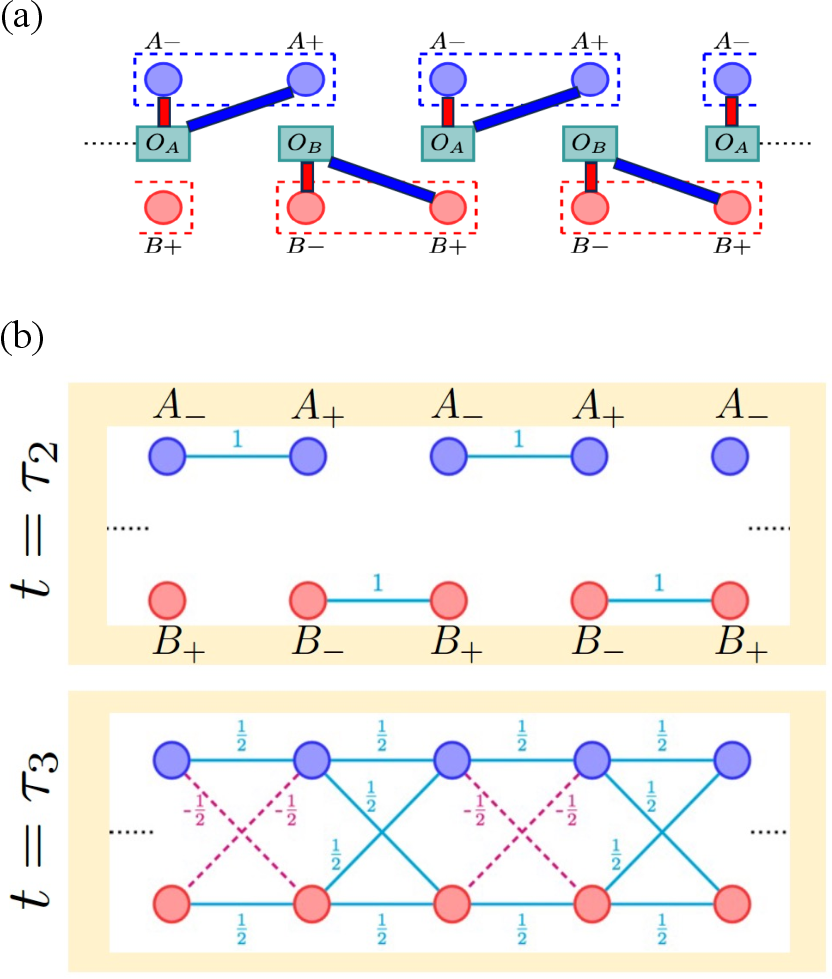}
    \caption{(a) Quantum operations in Step~II, where a beam-splitter interaction (red) and a parametric-amplification interaction (blue) are applied simultaneously.
    (b) Graph representations of the adjacency matrices at times $\tau_2$ and $\tau_3$. The edge weights shown in the graph correspond to adjacency-matrix elements divided by the common factor $\tanh(2r)$.}
    \label{fig2}
\end{figure}
\subsection*{Step II}
In Step II, we generate two-mode squeezing operations between pairs of mechanical modes. Under the Gaussian dynamics considered here, these operations create entanglement between the coupled modes.  Specifically, we couple the upper mechanical modes $b_{n}$ and $b_{n+1}$ for odd $n$, and the lower mechanical modes $b_{N+n}$ and $b_{N+n+1}$ for even $n$ ($n \in [1, N]$). Although these mechanical modes are not directly coupled, each pair is simultaneously linked to a common cavity mode $a_n$ via radiation-pressure interactions. For the upper mechanical modes, the associated Hamiltonian takes the form~\cite{TianPRL2013}: 
\begin{equation}
    H_{\rm I} = g_1 \left( b_n^\dagger a_n + a_n^\dagger b_n \right) 
    + ig_2  \left( b_{n+1}^\dagger a_n^\dagger  - a_n b_{n+1} \right), \label{eq:step2_op}
\end{equation}
which describes a red-detuned drive coupling $b_n$ to $a_n$ and a blue-detuned drive coupling $b_{n+1}$ to $a_n$, as illustrated in Fig.~\ref{fig2}(a). Here, the coupling strengths are parametrized as $g_1 = g_0 \cosh(r)$ and $g_2 = - g_0 \sinh(r)$, with $g_0$ the effective coupling rate and $r$ the two-mode squeezing parameter for this protocol. Details of the operator evolution under Eq.~\eqref{eq:step2_op} are provided in Appendix~\ref{appendix:a}.
Applying this Hamiltonian for a duration $\pi/g_0$, the mechanical and cavity modes become completely decoupled at the end of the evolution. At the time $\tau_2 = \tau_1 + \pi/g_0$, the quadratures become
\begin{subequations}
    \begin{align}
        \!\!\begin{pmatrix}
            Q_n(\tau_2) \\ P_{n+1}(\tau_2) 
        \end{pmatrix} \!\!=\!\!
        \begin{pmatrix}
            -\cosh(2r) & \sinh(2r) \\
            -\sinh(2r) & \cosh(2r) 
        \end{pmatrix} \!\!
        \begin{pmatrix}
            Q_n(\tau_1) \\ P_{n+1}(\tau_1)
        \end{pmatrix}, \\
        \!\!\begin{pmatrix}
            Q_{n+1}(\tau_2) \\ P_n(\tau_2) 
        \end{pmatrix}\!\!=\!\!
        \begin{pmatrix}
            \cosh(2r) & -\sinh(2r) \\
            \sinh(2r) & -\cosh(2r) 
        \end{pmatrix} \!\!
        \begin{pmatrix}
            Q_{n+1}(\tau_1) \\ P_n(\tau_1)
        \end{pmatrix}.
    \end{align}
\end{subequations}
As shown in Appendix~\ref{appendix:a}, this operation generates two-mode squeezing directly between the mechanical modes $b_n$ and $b_{n+1}$, thereby establishing entanglement between previously uncoupled modes.

\subsection*{Step III}
The final step of the protocol implements a beam-splitter operation between the mechanical modes $b_n$ and $b_{N+n}$ for all $n \in [1, N]$ by coupling both modes to the same cavity mode. A one-step realization of this effective beam-splitter interaction is achieved with the Hamiltonian
\begin{equation}     
    H_{\rm I} = g_1 \left( b_n^\dagger a_n + a_n^\dagger b_n \right) 
    + g_2 \left( b_{N+n}^\dagger a_n + a_n^\dagger b_{N+n} \right),
    \label{eq:step3_op}
\end{equation}
where the coupling strengths are parametrized as $g_1 = g_0 \sin \theta$ and $g_2 = g_0 \cos \theta$, with $g_0$ the overall coupling rate and $\theta$ an angle that controls the ratio between $g_1$ and $g_2$. This Hamiltonian corresponds to simultaneous red-detuned drives coupling $a_n$ to both $b_n$ and $b_{N+n}$. Applying this interaction for a duration $\pi/g_0$ with $\theta = \pi/8$ yields the quadrature transformations  
\begin{subequations}
    \begin{align}
        \begin{pmatrix}
            Q_n(\tau_3) \\
            Q_{N+n}(\tau_3)
        \end{pmatrix} &=
        \begin{pmatrix}
            \tfrac{1}{\sqrt{2}} & -\tfrac{1}{\sqrt{2}} \\
            -\tfrac{1}{\sqrt{2}} & -\tfrac{1}{\sqrt{2}}
        \end{pmatrix}
        \begin{pmatrix}
            Q_n(\tau_2) \\
            Q_{N+n}(\tau_2)
        \end{pmatrix}, \\
        \begin{pmatrix}
            P_n(\tau_3) \\
            P_{N+n}(\tau_3)
        \end{pmatrix} &=
        \begin{pmatrix}
            \tfrac{1}{\sqrt{2}} & -\tfrac{1}{\sqrt{2}} \\
            -\tfrac{1}{\sqrt{2}} & -\tfrac{1}{\sqrt{2}}
        \end{pmatrix}
        \begin{pmatrix}
            P_n(\tau_2) \\
            P_{N+n}(\tau_2)
        \end{pmatrix},
    \end{align}
\end{subequations}
where $\tau_3 = \tau_2 + \pi/g_0$. This transformation realizes a 50:50 beam-splitter between the two mechanical modes, with the cavity mode acting only as a mediator and becoming completely decoupled at the end of the evolution. A detailed derivation is provided in Appendix~\ref{appendix:b}.

\subsection*{Symplectic Matrix}
We define the $4N$-dimensional mechanical quadrature vector $\mathbf{R} = (Q_1, Q_2, \cdots, Q_{2N}, P_1, P_2, \cdots, P_{2N})^T$. The unitary operations described above can be represented by symplectic transformations acting on the quadrature vector $\mathbf{R}$~\cite{SimonPRL2000}. After Step~I, the quadrature vector transforms as $\mathbf{R}(\tau_1) = \mathbf{S_1}\mathbf{R}(0)$ with the corresponding symplectic matrix
\begin{equation}
    \mathbf{S_1} = 
    \begin{pmatrix} 
        e^\alpha \mathbf{I}_{2N} & \mathbf{0}_{2N} \\
        \mathbf{0}_{2N} & e^{-\alpha} \mathbf{I}_{2N}
    \end{pmatrix} \mathbf{U_0},
\end{equation}
where $\mathbf{I}_{2N}$ ($\mathbf{0}_{2N}$) denotes the $2N$-dimensional identity (zero) matrix, and $\mathbf{U_0}$ is a rotation acting on the $4N$ quadratures defined by
\begin{subequations}
    \begin{align}
       &Q_n \rightarrow \frac{Q_n + P_{N+n}}{\sqrt{2}}, \,\, P_{N+n} \rightarrow \frac{-Q_n + P_{N+n}}{\sqrt{2}}, \\ 
       &Q_{N+n} \rightarrow \frac{Q_{N+n} + P_{n}}{\sqrt{2}},\,\, P_{n} \rightarrow \frac{-Q_{N+n} + P_{n}}{\sqrt{2}}.
    \end{align}
\end{subequations}
Thus, Step I corresponds to a squeezing operation applied to rotated quadratures. Let the symplectic matrices corresponding to Step II and Step III be $\mathbf{S_2}$ and $\mathbf{S_3}$, respectively. After Step II, the quadrature vector becomes $\mathbf{R}(\tau_2) = \mathbf{S_2}\mathbf{S_1}\mathbf{R}(0)$. The total symplectic matrix for the three-step protocol is given by $\mathbf{S} = \mathbf{S_3}\mathbf{S_2}\mathbf{S_1}$, such that $\mathbf{R}(\tau_3) = \mathbf{S}\mathbf{R}(0)$. The explicit forms of these matrices can be obtained directly from the transformations derived above. 
A general symplectic matrix can be written as
\begin{equation}
    \mathbf{S} =
    \begin{pmatrix}
        \mathbf{S_A} & \mathbf{S_B} \\
        \mathbf{S_C} & \mathbf{S_D}
    \end{pmatrix},
\end{equation}
where $\mathbf{S_A}$, $\mathbf{S_B}$, $\mathbf{S_C}$, and $\mathbf{S_D}$ are $2N\times 2N$ matrices. The explicit expressions for these matrix blocks at the final time $\tau_3$ are given in Appendix~\ref{appendix:c} for the case of $N=3$ (six mechanical modes).

\subsection*{Adjacency Matrix}
We split the quadrature vector into displacement and momentum components with $\mathbf{R}^T = (\mathbf{Q}^T, \mathbf{P}^T)$. For a given symplectic matrix $\mathbf{S}$ at time $t$, the quadratures transform as
\begin{subequations} \label{eq:Rt}
    \begin{align}
        \mathbf{Q}(t) &= \mathbf{S_A} \mathbf{Q}(0) + \mathbf{S_B}\mathbf{P}(0), \\ 
        \mathbf{P}(t) &= \mathbf{S_C} \mathbf{Q}(0) + \mathbf{S_D}\mathbf{P}(0).
    \end{align} 
\end{subequations}
The adjacency matrix $\mathbf{A}$ associated with $\mathbf{S}$ is a $2N \times 2N$ matrix representing a graph with $2N$ vertices. The corresponding $2N$ nullifiers are defined by $\mathbf{N} = \mathbf{P} - \mathbf{A}\mathbf{Q}$. Using Eq.~(\ref{eq:Rt}), we obtain
\begin{equation}
    \mathbf{N}(t) = (\mathbf{S_C} - \mathbf{A}\mathbf{S_A})\mathbf{Q}(0)
    + (\mathbf{S_D} - \mathbf{A}\mathbf{S_B})\mathbf{P}(0). \label{eq:Nt}
\end{equation}
For an approximate CV cluster state, the covariance matrix ${\rm cov}(\mathbf{N}(t))$ should vanish in the limit of infinite momentum squeezing, $\alpha\rightarrow\infty$, analogous to the canonical construction of CV cluster states. The adjacency matrix can therefore be obtained as a function of the two-mode squeezing parameter $r$ introduced in Step~II:
\begin{equation}
    \mathbf{A}(r) = \lim_{\alpha \to \infty} \mathbf{S_C}\mathbf{S_A}^{-1} 
              = \lim_{\alpha\to \infty} \mathbf{S_D}\mathbf{S_B}^{-1}.  
\end{equation}

Using this procedure, we derive the adjacency matrices after each step of the protocol. The graphs in Fig.~\ref{fig2}(b) correspond to the adjacency matrices obtained after Steps II and III. For a quantum network with $N=3$ (six mechanical modes), the adjacency matrix after Step III is given by $\mathbf{A}(r)=\tanh(2r)\mathbf{A_0}$, where
\begin{equation}
    \mathbf{A_0} = 
    \begin{pmatrix}
        0 & \tfrac{1}{2} & 0 & 0 & -\tfrac{1}{2} & 0 \\
        \tfrac{1}{2} & 0 & \tfrac{1}{2} & -\tfrac{1}{2} & 0 & \tfrac{1}{2} \\
        0 & \tfrac{1}{2} & 0 & 0 & \tfrac{1}{2} & 0 \\
        0 & -\tfrac{1}{2} & 0 & 0 & \tfrac{1}{2} & 0 \\
        -\tfrac{1}{2} & 0 & \tfrac{1}{2} & \tfrac{1}{2} & 0 & \tfrac{1}{2} \\
        0 & \tfrac{1}{2} & 0 & 0 & \tfrac{1}{2} & 0
    \end{pmatrix}. \label{eq:A6modes}
\end{equation}
Here, all edge weights are uniformly scaled by the factor $\tanh(2r)$. The nullifiers for this six-mode system can therefore be written as, for example, $N_1 = P_1 - \tfrac{1}{2}\tanh(2r)(Q_2 - Q_5)$, $N_2 = P_2 - \tfrac{1}{2}\tanh(2r)(Q_1 + Q_3 - Q_4 + Q_6)$, and so on.

For completeness, we also consider the general adjacency matrix $\mathbf{A}(r,\alpha) = \mathbf{S_C}\mathbf{S_A}^{-1}$, as well as the limiting form $\mathbf{A_0}$ obtained from $\mathbf{A}(r)$ in the limit $r\to\infty$. Their properties and the associated nullifiers are discussed in Appendix~\ref{appendix:d} and Appendix~\ref{appendix:e}.

\begin{figure}
    \centering
    \includegraphics[width=8.2cm, clip]{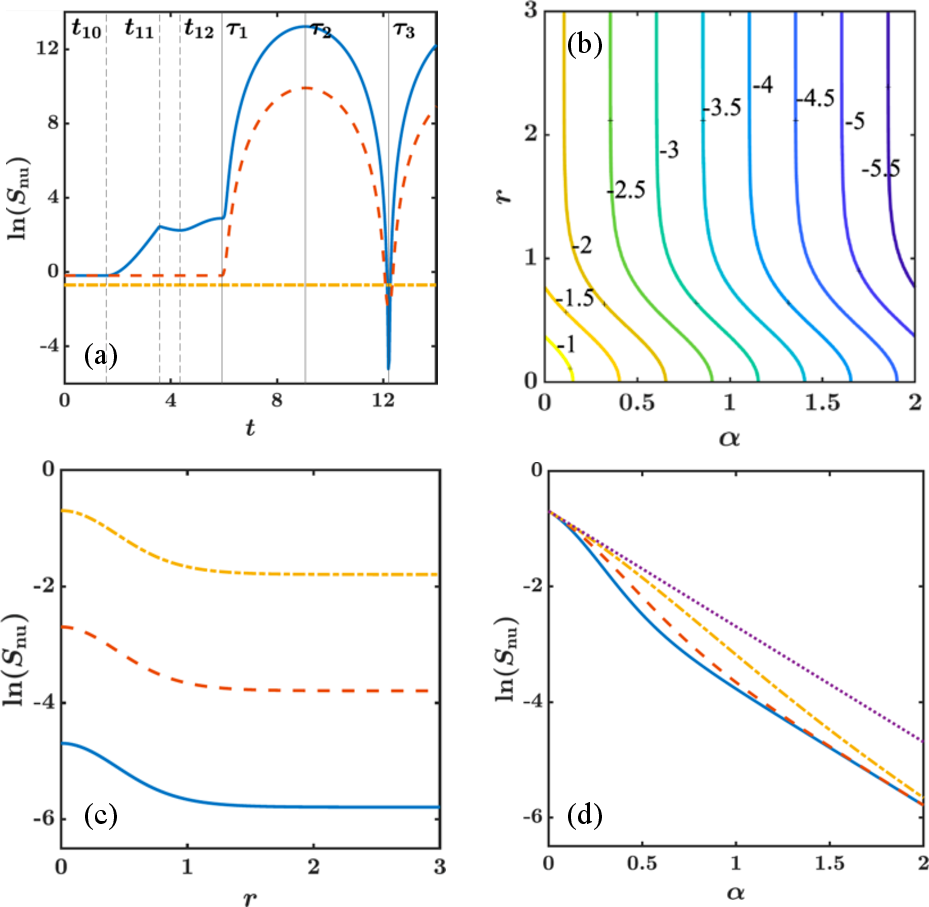}
    \caption{(a) Logarithm of the average nullifier $\ln(S_{\rm nu})$ vs time $t$. The squeezing parameters $(r,\alpha)=(3,2)$, $(3,0)$, and $(0,0)$ correspond to the solid, dashed, and dotted curves, respectively. Vertical lines indicate the times at which the three protocol steps are completed.
    (b) Contour plot of $\ln(S_{\rm nu})$ vs $r$ and $\alpha$.
    (c) $\ln(S_{\rm nu})$ vs $r$ for $\alpha=2,1,0$ (solid, dashed, and dash-dotted curves).
    (d) $\ln(S_{\rm nu})$ vs $\alpha$ for $r/\alpha=1.5,1,0.5,0$ (solid, dashed, dash-dotted, and dotted curves). Here $g_0=1$ sets the dimensionless energy unit, and all plots are for a network with $N=3$ (six mechanical modes)}.
    \label{fig3}
\end{figure}
\subsection*{Average Nullifier}
To characterize the CV cluster states generated by our protocol, we evaluate the nullifier variances ${\rm cov}(N_n)$ at the final time $\tau_3$. We define $\mathbf{C_N}=\langle \mathbf{N}(\tau_3) \mathbf{N}^T(\tau_3)\rangle$. Assuming that the mechanical modes are initially prepared in thermal states with thermal phonon number $n_{\rm th}$, the quadrature correlations at $t=0$ are given by
$\langle Q_m(0)Q_n(0)\rangle = \langle P_m(0)P_n(0)\rangle = \delta_{mn}\!\left(n_{\rm th}+\tfrac{1}{2}\right)$ and $\langle Q_m(0)P_n(0)\rangle = -\langle P_n(0)Q_m(0)\rangle = \tfrac{i}{2}\delta_{mn}$. Using Eq.~(\ref{eq:Nt}), we obtain $\mathbf{C_N} = (n_{\rm th}+\frac{1}{2}) \mathbf{C_d}$, with
\begin{align}
    \mathbf{C_d} &= (\mathbf{S_C}-\mathbf{A}\mathbf{S_A})(\mathbf{S_C}-\mathbf{A}\mathbf{S_A})^T \nonumber \\
    &\quad + (\mathbf{S_D}-\mathbf{A}\mathbf{S_B})(\mathbf{S_D}-\mathbf{A}\mathbf{S_B})^T.
\end{align}
The diagonal elements of $\mathbf{C_N}$ give the variances of the individual nullifiers, ${\rm cov}(N_n)$. Their average is given by $S_{\rm nu} = \tfrac{1}{2N}{\rm Tr}[\mathbf{C_N}]$. Analytical expressions for $S_{\rm nu}$ can be derived for arbitrary system size $N$ following the same procedure. The results for $N=4$--$6$ are presented in Appendix~\ref{appendix:e}.

For the six-mode mechanical network ($N=3$) at zero temperature ($n_{\rm th}=0$), we obtain
\begin{equation}
    S_{\rm nu} = \frac{e^{-2\alpha}}{6} \left[ 1 + 2\sech^2{(2r)}\right], \label{eq:Nu_ave}
\end{equation}
which depends on both the initial momentum squeezing parameter $\alpha$ introduced in Step~I and the two-mode squeezing parameter $r$ introduced in Step~II. In the limit $\alpha\to\infty$, one has $S_{\rm nu}\to 0$, corresponding to an approximate CV cluster state. To confirm this analytical result, we numerically simulate the Gaussian-state dynamics of the protocol and calculate the average nullifier using the adjacency matrix $\mathbf{A}(r)$. 
Figure~\ref{fig3}(a) shows the time evolution of $\ln(S_{\rm nu})$, where the beam-splitter operation in Step~III is continued beyond $\tau_3$. Figure~\ref{fig3}(b) presents a contour plot of $\ln(S_{\rm nu})$ as a function of $r$ and $\alpha$. The average nullifier decreases with increasing squeezing parameters, confirming the generation of approximate CV cluster states. Furthermore, Fig.~\ref{fig3}(c) and (d) show $\ln(S_{\rm nu})$ as a function of $r$ and $\alpha$, respectively. The numerical results demonstrate that the average nullifier decreases monotonically with both $r$ and $\alpha$, indicating that the quality of the generated CV cluster states can be improved by increasing the squeezing parameters in Steps~I and II.

\section{Damping and Scalability \label{sec:decoherence}}
In a practical phononic quantum network, both the cavity and the mechanical modes are subject to finite damping and thermal fluctuations. In this section, we analyze the performance of the proposed protocol in the presence of cavity and mechanical dissipation.
To describe the dynamics of the phononic quantum network, we introduce a super-quadrature vector that incorporates both mechanical and cavity quadratures: $\mathbf{\widetilde{R}}^T=(\mathbf{\widetilde{Q}}^T,\mathbf{\widetilde{P}}^T)$ with $\mathbf{\widetilde{Q}}^T=(\mathbf{Q}^T,Q_{a1},\cdots,Q_{aN})$ and $\mathbf{\widetilde{P}}^T=(\mathbf{P}^T,P_{a1},\cdots,P_{aN})$. For an array of $N$ mechanical resonators containing $2N$ mechanical modes and $N$ cavity modes, $\mathbf{\widetilde{R}}$ is a $6N$-dimensional vector. The density matrix $\rho$ of the quantum network evolves according to the master equation $\frac{d\rho}{dt}=-i[H_{\rm I},\rho]+\sum_m \mathcal{L}_m[\rho]$,
where the interaction Hamiltonian is $H_{\rm I}=\frac{1}{2} \mathbf{\widetilde{R}}^T \mathbf{G}\mathbf{\widetilde{R}}$ with $\mathbf{G}$ a symmetric matrix, and $\mathcal{L}_m$ is the Liouvillian superoperator describing the damping of the $m$th mode (see Appendix~\ref{appendix:f}).

Any Gaussian state is completely characterized by its first- and second-order moments~\cite{WeedbrookRMP2012GaussianQI}. In our system, the first-order moments vanish, and the state is therefore fully specified by the covariance matrix $\mathbf{V}$ of $\mathbf{\widetilde{R}}$, with matrix elements $V_{mn}=\tfrac{1}{2}\langle \widetilde{R}_m \widetilde{R}_n+\widetilde{R}_n \widetilde{R}_m\rangle$. Using the above master equation, the covariance matrix evolves according to~\cite{KogaPRA2012}
\begin{equation} 
    \frac{d\mathbf{V}}{dt} = \mathbf{B} \mathbf{V} + \mathbf{V} \mathbf{B}^{T} + \mathbf{D}, \label{eq:dVdt}
\end{equation}
with $\mathbf{B} = \mathbf{\Sigma}(\mathbf{G}  + {\rm Im}[\mathbf{C}^{\dagger}\mathbf{C}])$, $\mathbf{D} = \mathbf{\Sigma} {\rm Re}[\mathbf{C}^{\dagger} \mathbf{C}] \mathbf{\Sigma}^{T}$, and  
\begin{equation}
    \mathbf{\Sigma} = \begin{pmatrix}
    0 & \mathbf{I_{3N}} \\
    -\mathbf{I_{3N}} & 0
    \end{pmatrix}.
\end{equation}
Here $\mathbf{C}$ is determined by the damping channels (Appendix~\ref{appendix:f}). We assume that the quantum network is initially uncorrelated, with the cavity modes in their ground states and the mechanical modes in thermal states with thermal occupation number $n_{\rm th}$. The initial quadrature variances are thus ${\rm cov}(Q_n)={\rm cov}(P_n)=n_{\rm th}+1/2$ and ${\rm cov}(Q_{an})={\rm cov}(P_{an})=1/2$. Consequently, the initial covariance matrix $\mathbf{V}(0)$ is block diagonal.

\begin{figure}
    \centering
    \includegraphics[width=8.5cm, clip]{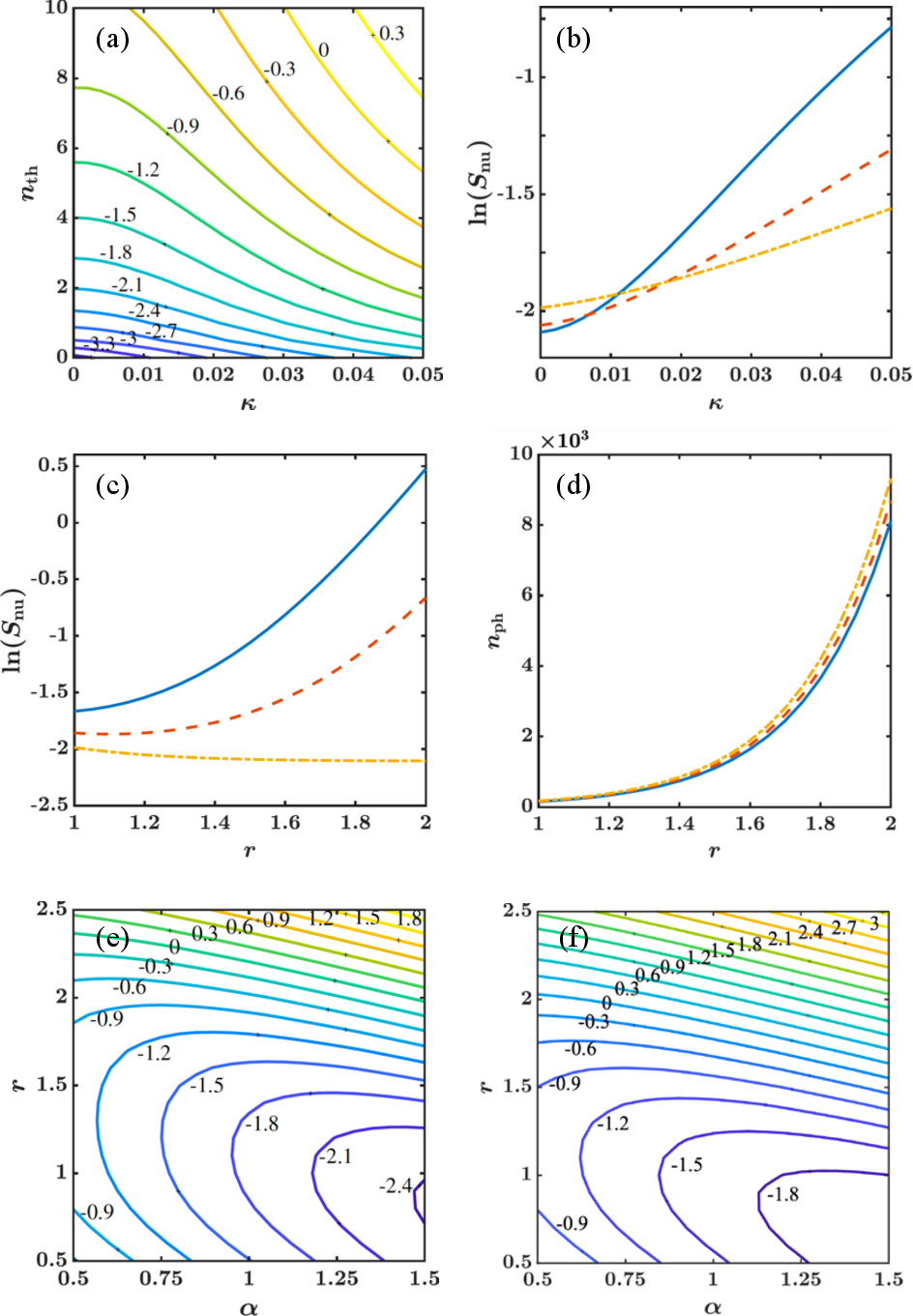}
    \caption{(a) Contour plot of $\ln(S_{\rm nu})$ vs $\kappa$ and $n_{\rm th}$ for $(r,\alpha)=(1.5,1)$.
    (b) $\ln(S_{\rm nu})$ vs $\kappa$ for $r=1.5,\,1.25,\,1$ (solid, dashed, and dash-dotted curves) with $\alpha=1$ and $n_{\rm th}=2$.
    (c) $\ln(S_{\rm nu})$ and (d) $n_{\rm ph}$ vs $r$ for $\kappa=0.04,\,0.02,\,0$ (solid, dashed, and dash-dotted curves) with $\alpha=1$ and $n_{\rm th}=2$.
    (e) and (f) Contour plots of $\ln(S_{\rm nu})$ vs $\alpha$ and $r$ for $\kappa=0.02$ and $0.04$, respectively, with $n_{\rm th}=2$. In all plots, $\gamma=5\times10^{-4}$ and $N=3$ (six mechanical modes).}
    \label{fig4}
\end{figure}
\subsection*{Effect of Damping}
Using Eq.~(\ref{eq:dVdt}), we simulate the dynamics of the covariance matrix and compute the logarithm of the average nullifier $\ln(S_{\rm nu})$ at the final time $\tau_3$ for a quantum network with $N=3$ (six mechanical modes). Figure~\ref{fig4}(a) shows a contour plot of $\ln(S_{\rm nu})$ versus the cavity damping rate $\kappa$ and the thermal occupation number $n_{\rm th}$ for $(r,\alpha)=(1.5,1)$. As expected, larger values of $\kappa$ and $n_{\rm th}$ degrade the cluster state. However, for moderate parameters $\kappa\sim0.02$, $\gamma=5\times10^{-4}$, and $n_{\rm th}=2$, we still obtain $\ln(S_{\rm nu})\sim-1.7$, demonstrating robust cluster-state generation.
Figure~\ref{fig4}(b) plots $\ln(S_{\rm nu})$ as a function of $\kappa$ for several values of $r$. Interestingly, at large $\kappa$, the average nullifier for $r=1.5$ becomes larger than those for $r=1$ and $1.25$, deviating from the analytical behavior shown in Fig.~\ref{fig3}(c) for zero damping. This effect is further illustrated in Fig.~\ref{fig4}(c), where $\ln(S_{\rm nu})$ may either increase or decrease with increasing $r$. These results show that damping can partially offset the benefit of stronger squeezing.
To understand this behavior, Fig.~\ref{fig4}(d) plots the average phonon number $n_{\rm ph}=\tfrac{1}{2N}\sum_n \langle b_n^\dagger b_n\rangle$ as a function of $r$. The phonon number increases rapidly with squeezing and exceeds $8\times10^3$ at $r=2$. Such highly excited states are more susceptible to damping, explaining why moderate squeezing becomes optimal in the presence of dissipation.
Figures~\ref{fig4}(e) and \ref{fig4}(f) show contour plots of $\ln(S_{\rm nu})$ as a function of $(r,\alpha)$ for $\kappa=0.02$ and $0.04$, respectively. In both cases, optimal performance is achieved at moderate squeezing, and the optimum shifts toward smaller values of $(r,\alpha)$ as the damping rate increases. 

The above results show that the proposed pulsed protocol remains effective in the presence of damping and thermal noise. Even under moderate decoherence, strongly correlated CV cluster states can be generated (see Sec.~\ref{sec:entanglementtransfer}), highlighting the potential of this approach for scalable CV quantum information processing in phononic quantum networks.
At the same time, the pulsed generation of CV cluster states requires optomechanical systems to operate in the strong-coupling regime, which may present experimental challenges. For example, while nanostructures such as optomechanical crystals can achieve relatively large optomechanical coupling rates, their optical decay rates often remain high. Promising alternatives include hybrid platforms based on collective excitations, such as magnons, which can simultaneously exhibit long coherence times and strong coupling to mechanical modes. Continued advances in these platforms may broaden the accessible operating regime of the protocol and further enhance the feasibility of generating and distributing entanglement in phononic quantum networks.

\begin{figure}
    \centering
    \includegraphics[width=8.5cm, clip]{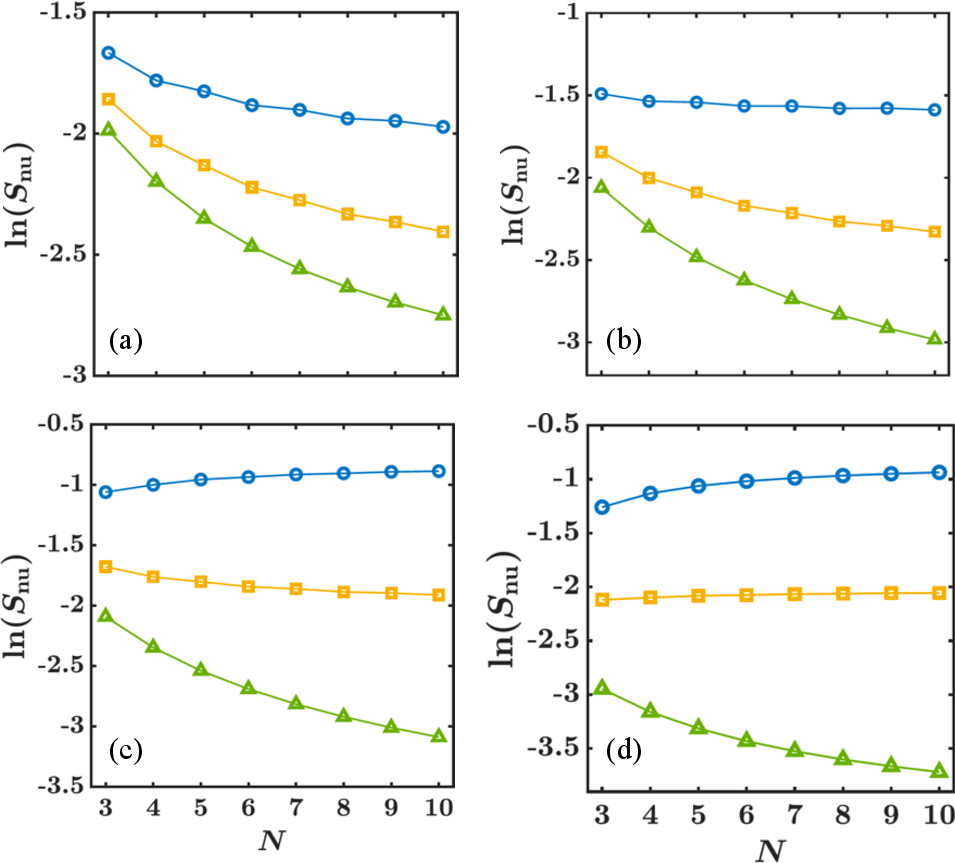}
    \caption{Logarithm of the average nullifier $\ln(S_{\rm nu})$ vs the size $N$ of a quantum network. The squeezing parameters $(r,\alpha)$ are (a) $(1,1)$, (b) $(1.25,1)$, (c) $(1.5,1)$, and (d) $(1.25,1.5)$. Here the cavity damping rates are $\kappa=0.04$, $0.02$, and $0$ (circles, squares, and triangles, respectively). In all panels, $\gamma=5\times10^{-4}$ and $n_{\rm th}=2$.}
    \label{fig5}
\end{figure}
\subsection*{Scalability}
In the proposed protocol, the number of quantum operations (control resources) scales linearly with the size $N$ of the quantum network. Another important question is whether this scalability can be maintained in the presence of dissipation and thermal noise. To address this issue, we extend the numerical simulations of the pulsed protocol to system sizes up to $N=10$, corresponding to $20$ mechanical modes and $10$ cavity modes. The results are presented in Fig.~\ref{fig5}, where the logarithm of the average nullifier, $\ln(S_{\rm nu})$, is plotted as a function of $N$ for three cavity damping rates and several representative choices of the squeezing parameters $(r,\alpha)$.
We find that for $\kappa=0$ and $0.02$ (zero or weak cavity damping), $\ln(S_{\rm nu})$ decreases with increasing $N$ in all four panels, indicating that the performance of the protocol improves as the network size grows. For stronger damping, $\kappa=0.04$, this behavior is preserved for moderate squeezing parameters, while a modest increase in $\ln(S_{\rm nu})$ with $N$ emerges at larger squeezing values. Overall, these results demonstrate that the proposed scheme remains scalable in the presence of moderate dissipation and squeezing.

\section{Entanglement of Distant Modes \label{sec:entanglementtransfer}}
\begin{figure}
    \centering
    \includegraphics[width=8.2cm, clip]{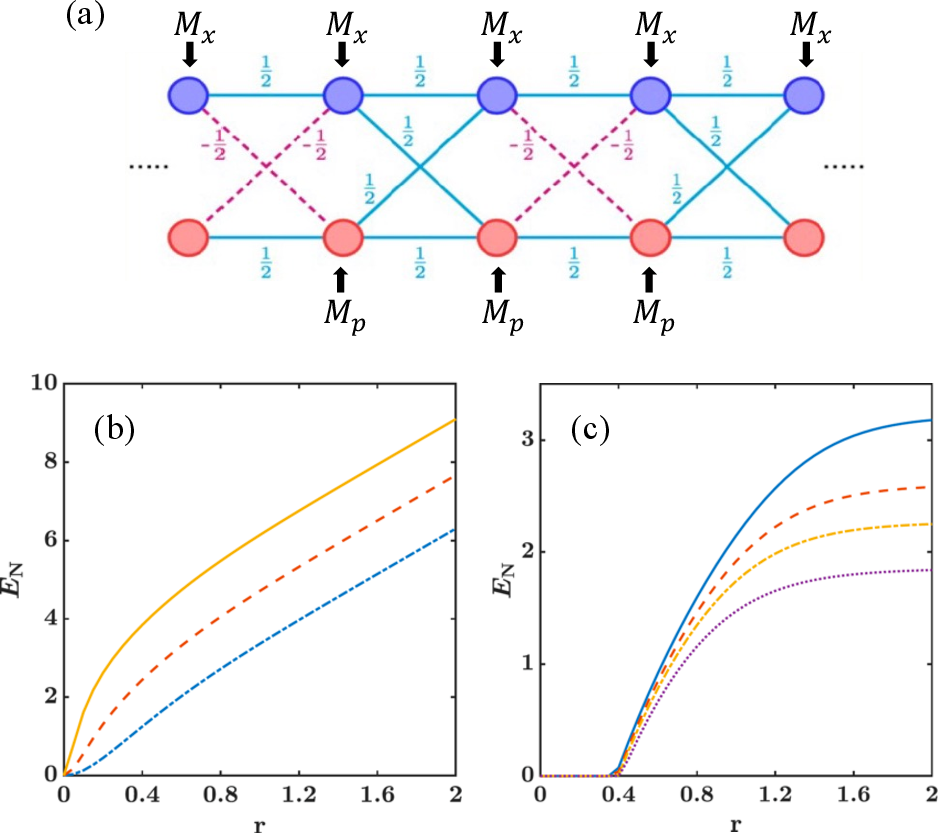}
    \caption{(a) Schematic of entanglement generation between distant mechanical modes through local quadrature measurements.
    (b) Logarithmic negativity $E_{\rm N}$ vs $r$ in the absence of damping. The solid, dashed, and dash-dotted curves correspond to $\alpha=1.5$, $1$, and $0.5$, respectively. 
    (c) $E_{\rm N}$ vs $r$ in the presence of damping. The solid, dashed, dash-dotted, and dotted curves correspond to $\kappa=0$, $0.01$, $0.02$, and $0.04$, respectively. Here $\alpha=1$, $\gamma=5\times10^{-4}$, and $n_{\rm th}=2$.}
    \label{fig6}
\end{figure}

CV cluster states can be used to generate entanglement between distant CV modes through quadrature measurements on other modes in the network~\cite{WeedbrookRMP2012GaussianQI, GuPRA2009}. A displacement measurement $M_x$ on a selected mode removes that mode from the cluster. In contrast, a momentum measurement $M_p$ removes the measured mode while preserving the correlations between its neighboring modes. We consider the measurement scheme illustrated in Fig.~\ref{fig6}(a) for a quantum network of $N$ mechanical resonators ($2N$ mechanical modes). In this protocol, displacement measurements are performed on the upper mechanical modes $b_n$ ($n\in[1,N]$), while momentum measurements are performed on the lower mechanical modes $b_m$ ($m\in[N+2,\,(2N-1)]$). These measurements remove the upper modes from the cluster while preserving the correlations between the two remaining modes, $b_{N+1}$ and $b_{2N}$, thereby generating entanglement between them~\cite{ZhangPRA2006}.
The covariance matrix of the modes $b_{N+1}$ and $b_{2N}$ after the measurements can be derived following the procedure outlined in Appendix~\ref{appendix:g}. From the resulting covariance matrix, we evaluate the logarithmic negativity $E_{\rm N}$ as a quantitative measure of the entanglement between the two modes~\cite{PlenioPRL2005, VidalPRA2002}.

Figure~\ref{fig6}(b) shows $E_{\rm N}$ as a function of the squeezing parameter $r$ for several values of $\alpha$ in the absence of damping. The entanglement increases with both $r$ and $\alpha$, reaching appreciable values even at moderate squeezing. Figure~\ref{fig6}(c) presents $E_{\rm N}$ as a function of $r$ in the presence of damping and thermal excitations for $\alpha=1$. Although damping reduces the magnitude of the entanglement, moderate values of $r$ and $\alpha$ still yield substantial entanglement. At finite temperature, however, small values of $r$ are insufficient to generate entanglement. These results demonstrate that distant mechanical modes in a phononic quantum network can be entangled through local quadrature measurements. As shown in previous studies~\cite{PaternostroPRA2009, WSonJModOptics2002}, such mechanical entanglement can be transferred to distant qubits that are not directly coupled. The CV cluster states generated in phononic quantum networks can therefore serve as a powerful resource for creating and distributing entanglement across remote quantum nodes.

\section{Conclusions \label{sec:conclusions}}
We have developed a scalable pulsed protocol for generating CV cluster states in phononic quantum networks. Using the symplectic matrix of the protocol, we derived the corresponding adjacency matrix and the average nullifier of the resulting cluster states, showing that the average nullifier approaches zero with increasing squeezing. We further showed that the protocol requires only $O(N)$ control resources and remains scalable in the presence of moderate dissipation and thermal noise. We analyzed the effects of damping and found that mechanical and cavity losses degrade cluster states at strong squeezing due to the large phonon population. As a result, the average nullifier exhibits a nonmonotonic dependence on the squeezing parameters, and optimal performance is achieved at moderate squeezing. In addition, we demonstrated that the entanglement contained in a CV cluster state can be transferred to two distant mechanical modes through local quadrature measurements, enabling long-range quantum correlations across the network.
Our protocol is applicable to a broad class of systems operating in the strong-coupling regime, where the light--matter interaction exceeds the cavity bandwidth. Compared with photonic platforms, mechanical CV cluster states offer long coherence times and can be generated in large-scale networks. These features make phononic quantum networks a promising platform for hybrid quantum information processing, supporting the generation, storage, and distribution of quantum entanglement.

\section*{Acknowledgements} L.T. was supported by NSF Awards No. DMR-2037987 and No. CICI-2530705, and the UC-MRPI Program (Grant No. M23PL5936). A.G. was also supported by the Sandbox AQ Fellowship. H.W was
supported by NSF Award No. 2514515.
\appendix

\section{One-step Generation of Two-mode Squeezing \label{appendix:a}}
In Step~II of our protocol, the Hamiltonian in Eq.~(\ref{eq:step2_op}) is applied to neighboring mechanical modes that are simultaneously coupled to a common cavity mode~\cite{TianPRL2013}. The corresponding operators evolve according to  
\begin{equation}  \label{eq:dOdt_step2}
    \frac{d}{dt}
    \begin{pmatrix}
        b_n \\[3pt] b_{n+1}^{\dagger} \\[3pt] a_n
    \end{pmatrix} =
    \begin{pmatrix}
        0 & 0 & -i g_1 \\
        0 & 0 & g_2 \\
        -i g_1 & g_2 & 0
    \end{pmatrix}
    \begin{pmatrix}
        b_n \\[3pt] b_{n+1}^{\dagger} \\[3pt] a_n
    \end{pmatrix}.
\end{equation}
Solving Eq.~(\ref{eq:dOdt_step2}) yields the time evolution of $b_n(t)$, $b_{n+1}^\dagger(t)$, and $a_n(t)$. In particular, after an interaction time $\Delta t = k\pi / g_0$, where $k$ is an odd integer, the mechanical (cavity) operators evolve independently of the cavity (mechanical) operators: 
\begin{equation}
    \begin{pmatrix}
        b_n(\Delta t) \\
        b_{n+1}^{\dagger}(\Delta t)
    \end{pmatrix} =
    \begin{pmatrix}
        -\cosh(2r) & i\sinh(2r) \\
        i\sinh(2r) & \cosh(2r)
    \end{pmatrix}
    \begin{pmatrix}
        b_n \\
        b_{n+1}^{\dagger}
    \end{pmatrix}.
\end{equation}
In terms of the quadrature operators, the evolution reads
\begin{subequations}
    \begin{align}
        Q_n(\Delta t) &= -\cosh(2r)\, Q_n + \sinh(2r)\, P_{n+1}, \\
        Q_{n+1}(\Delta t) &= \cosh(2r)\, Q_{n+1} - \sinh(2r)\, P_{n}, \\
        P_n(\Delta t) &= -\cosh(2r)\, P_n + \sinh(2r)\, Q_{n+1}, \\
        P_{n+1}(\Delta t) &= -\sinh(2r)\, Q_{n} + \cosh(2r)\, P_{n+1}.
    \end{align}
\end{subequations}
The resulting correlations between the two mechanical modes satisfy 
\begin{subequations}
    \begin{align}
        P_n(\Delta t) - Q_{n+1}(\Delta t) &= -e^{-2r}\,\big(P_n + Q_{n+1}\big), \\
        P_{n+1}(\Delta t) - Q_{n}(\Delta t) &= \;\; e^{-2r}\,\big(P_{n+1} + Q_{n}\big).
    \end{align}
\end{subequations}
Thus, in the limit $r \to \infty$, both $P_n(\Delta t) - Q_{n+1}(\Delta t)$ and $P_{n+1}(\Delta t) - Q_{n}(\Delta t)$ vanish, corresponding to an ideal two-mode squeezed state.

\medskip

\section{One-step Generation of Beam-splitter Operation \label{appendix:b}}
In Step~III, the Hamiltonian in Eq.~(\ref{eq:step3_op}) is applied to a pair of mechanical modes and their associated cavity mode. The time evolution of the operators is governed by
\begin{equation} \label{eq:dOdt_step3}
    \frac{d}{dt}
    \begin{pmatrix}
        b_n \\[2pt] b_{N+n} \\[2pt] a_n
    \end{pmatrix} =
    \begin{pmatrix}
        0 & 0 & -i g_1 \\
        0 & 0 & -i g_2 \\
        -i g_1 & -i g_2 & 0
    \end{pmatrix}
    \begin{pmatrix}
        b_n \\[2pt] b_{N+n} \\[2pt] a_n
    \end{pmatrix}.
\end{equation}
Solving Eq.~(\ref{eq:dOdt_step3}), we find that after an interaction time $\Delta t = k \pi / g_0$, where $k$ is an odd integer, the mechanical and cavity operators decouple. The mechanical operators evolve as
\begin{equation}
    \begin{pmatrix}
        b_n(\Delta t) \\
        b_{N+n}(\Delta t)
    \end{pmatrix} =
    \begin{pmatrix}
        \cos(2\theta) & -\sin(2\theta) \\
        -\sin(2\theta) & -\cos(2\theta)
    \end{pmatrix}
    \begin{pmatrix}
        b_n \\
        b_{N+n}
    \end{pmatrix}.
\end{equation}
The corresponding quadrature operators are
\begin{subequations}
    \begin{align}
        \begin{pmatrix}
            Q_n(\Delta t) \\
            Q_{N+n}(\Delta t)
        \end{pmatrix} &=
        \begin{pmatrix}
            \cos(2\theta) & -\sin(2\theta) \\
            -\sin(2\theta) & -\cos(2\theta)
        \end{pmatrix}
        \begin{pmatrix}
            Q_n \\
            Q_{N+n}
        \end{pmatrix}, \\
        \begin{pmatrix}
            P_n(\Delta t) \\
            P_{N+n}(\Delta t)
        \end{pmatrix} &=
        \begin{pmatrix}
            \cos(2\theta) & -\sin(2\theta) \\
            -\sin(2\theta) & -\cos(2\theta)
        \end{pmatrix}
        \begin{pmatrix}
            P_n \\
            P_{N+n}
        \end{pmatrix}.
    \end{align}
\end{subequations}
This transformation corresponds to a single-step beam-splitter operation between the mechanical modes $b_n$ and $b_{N+n}$. For $\theta=\pi/8$, it becomes a $50{:}50$ beam splitter. The cavity mode acts only as a mediating bus and becomes disentangled from the mechanical modes at the end of the interaction.

\section{Components of Symplectic Matrix \label{appendix:c}}
For a phononic quantum network with $N=3$ mechanical resonators (i.e., six mechanical modes), the components of the total symplectic matrix $\mathbf{S}$ can be derived explicitly as
\begin{widetext}
    \begin{equation}
    \mathbf{S_A} = \frac{1}{2}\begin{pmatrix}
        -\cosh(2r) e^{\alpha} & 0 & 0 & -e^{\alpha} & -\sinh(2r) e^{-\alpha} & 0 \\
 0 & \cosh(2r) e^{\alpha} & \sinh(2r) e^{-\alpha} & \sinh(2r) e^{-\alpha} & \cosh(2r) e^{\alpha} & 0 \\
 0 & -\sinh(2r) e^{-\alpha} & e^{\alpha} & 0 & 0 & -\cosh(2r) e^{\alpha} \\
 \cosh(2r) e^{\alpha} & 0 & 0 & -e^{\alpha} & \sinh(2r) e^{-\alpha} & 0 \\
 0 & -\cosh(2r) e^{\alpha} & \sinh(2r) e^{-\alpha} & -\sinh(2r) e^{-\alpha} & \cosh(2r) e^{\alpha} & 0 \\
 0 & -\sinh(2r) e^{-\alpha} & -e^{\alpha} & 0 & 0 & -\cosh(2r) e^{\alpha}
    \end{pmatrix},
\end{equation}
\begin{equation}
    \mathbf{S_B} = \frac{1}{2}\begin{pmatrix}
 -e^{\alpha} & \sinh(2r) e^{-\alpha} & 0 & -\cosh(2r) e^{\alpha} & 0 & 0 \\
 -\sinh(2r) e^{-\alpha} & \cosh(2r) e^{\alpha} & 0 & 0 & \cosh(2r) e^{\alpha} & -\sinh(2r) e^{-\alpha} \\
 0 & 0 & -\cosh(2r) e^{\alpha} & 0 & \sinh(2r) e^{-\alpha} & e^{\alpha} \\
 -e^{\alpha} & -\sinh(2r) e^{-\alpha} & 0 & \cosh(2r) e^{\alpha} & 0 & 0 \\
 \sinh(2r) e^{-\alpha} & \cosh(2r) e^{\alpha} & 0 & 0 & -\cosh(2r) e^{\alpha} & -\sinh(2r) e^{-\alpha} \\
 0 & 0 & -\cosh(2r) e^{\alpha} & 0 & \sinh(2r) e^{-\alpha} & -e^{\alpha}
\end{pmatrix},
\end{equation}
\begin{equation}
    \mathbf{S_C} = \frac{1}{2}\begin{pmatrix}
 e^{-\alpha} & \sinh(2r) e^{\alpha} & 0 & \cosh(2r) e^{-\alpha} & 0 & 0 \\
 -\sinh(2r) e^{\alpha} & -\cosh(2r) e^{-\alpha} & 0 & 0 & -\cosh(2r) e^{-\alpha} & -\sinh(2r) e^{\alpha} \\
 0 & 0 & \cosh(2r) e^{-\alpha} & 0 & \sinh(2r) e^{\alpha} & -e^{-\alpha} \\
 e^{-\alpha} & -\sinh(2r) e^{\alpha} & 0 & -\cosh(2r) e^{-\alpha} & 0 & 0 \\
 \sinh(2r) e^{\alpha} & -\cosh(2r) e^{-\alpha} & 0 & 0 & \cosh(2r) e^{-\alpha} & -\sinh(2r) e^{\alpha} \\
 0 & 0 & \cosh(2r) e^{-\alpha} & 0 & \sinh(2r) e^{\alpha} & e^{-\alpha}
\end{pmatrix},
\end{equation}
\begin{equation}
    \mathbf{S_D} = \frac{1}{2}\begin{pmatrix}
 -\cosh(2r) e^{-\alpha} & 0 & 0 & -e^{-\alpha} & \sinh(2r) e^{\alpha} & 0 \\
 0 & \cosh(2r) e^{-\alpha} & -\sinh(2r) e^{\alpha} & -\sinh(2r) e^{\alpha} & \cosh(2r) e^{-\alpha} & 0 \\
 0 & \sinh(2r) e^{\alpha} & e^{-\alpha} & 0 & 0 & -\cosh(2r) e^{-\alpha} \\
 \cosh(2r) e^{-\alpha} & 0 & 0 & -e^{-\alpha} & -\sinh(2r) e^{\alpha} & 0 \\
 0 & -\cosh(2r) e^{-\alpha} & -\sinh(2r) e^{\alpha} & \sinh(2r) e^{\alpha} & \cosh(2r) e^{-\alpha} & 0 \\
 0 & \sinh(2r) e^{\alpha} & -e^{-\alpha} & 0 & 0 & -\cosh(2r) e^{-\alpha}
\end{pmatrix}. 
\end{equation}
\end{widetext}

\section{Adjacency Matrix $\mathbf{A}(r,\alpha)$ and $\mathbf{A}_0$ \label{appendix:d}}
In Sec.~\ref{sec:protocol}, we derived the adjacency matrix by taking the limit of infinite initial momentum squeezing. Here, we consider the more general adjacency matrix
\begin{equation}
\mathbf{A}(r,\alpha)=\mathbf{S_C}\mathbf{S_A}^{-1},
\end{equation}
which depends on both the initial momentum squeezing parameter $\alpha$ introduced in Step~I and the two-mode squeezing parameter $r$ introduced in Step~II. A definition based on $\mathbf{A}(r,\alpha)=\mathbf{S_D}\mathbf{S_B}^{-1}$ can also be considered and yields similar results.
With this definition, the first term in Eq.~(\ref{eq:Nt}) vanishes, and the nullifier becomes $\mathbf{N}=(\mathbf{S_D}-\mathbf{A}\mathbf{S_B})\mathbf{P}(0)$, since $\mathbf{S_D}-\mathbf{A}\mathbf{S_B}\neq0$. For $N=3$, we obtain
\begin{equation}
 \mathbf{A}(r, \alpha)=
\begin{pmatrix}
A_{11} & A_{12} & A_{13} & 0 &
A_{15} & -A_{13} \\
A_{12} & A_{11} & -A_{15}  & A_{15} & 0 & A_{12} \\
A_{13} & -A_{15} & -A_{11} &
A_{13} & A_{12} & 0 \\
0 & A_{15} & A_{13} & -A_{11} &
A_{12} & -A_{13} \\
A_{15} & 0 & A_{12} & A_{12} & -A_{11} & -A_{15} \\
-A_{13} & A_{12} & 0 &
-A_{13} & -A_{15} & A_{11}
\end{pmatrix}, \label{eq:A6modes_r_alpha}
\end{equation}
with the coefficients
\begin{subequations}
    \begin{align}
        A_{11} & = -e^{-2\alpha}\sech(2r), \\
        A_{12} & = \frac{1}{2}\left(1-e^{-4\alpha}\sech(2r)\right)\tanh(2r), \\
        A_{13} & = \frac{1}{2}e^{-6\alpha}\tanh^2(2r), \\
        A_{15} & = -\frac{1}{2}\left(1+e^{-4\alpha}\sech(2r)\right)\tanh(2r).
    \end{align}
\end{subequations}
Compared with $\mathbf{A}(r)$ in Sec.~\ref{sec:protocol}, the adjacency matrix $\mathbf{A}(r,\alpha)$ contains additional weak links with weights $\pm A_{13}$, as well as nonzero diagonal (self-weight) elements $\pm A_{11}$. The edge weights $\pm A_{12}$ and $\pm A_{15}$ acquire small corrections of order $O(e^{-4\alpha})$. In the limit $\alpha\rightarrow\infty$, all of these corrections vanish, and $\mathbf{A}(r,\alpha)$ reduces to $\mathbf{A}(r)$.

In the limit $r \rightarrow\infty$, the adjacency matrix becomes
\begin{equation}
 \lim_{r\rightarrow\infty} \mathbf{A}(r, \alpha)=
\begin{pmatrix}
0 & \frac{1}{2} & A_{13}^{\infty} & 0 & -\frac{1}{2} & -A_{13}^{\infty} \\
\frac{1}{2} & 0 & \frac{1}{2}  & -\frac{1}{2} & 0 & \frac{1}{2} \\
A_{13}^{\infty} & \frac{1}{2} & 0 & A_{13}^{\infty} & \frac{1}{2} & 0 \\
0 & -\frac{1}{2} & A_{13}^{\infty} & 0 & \frac{1}{2} & -A_{13}^{\infty} \\
-\frac{1}{2} & 0 & \frac{1}{2} & \frac{1}{2} & 0 & \frac{1}{2} \\
-A_{13}^{\infty} & \frac{1}{2} & 0 & -A_{13}^{\infty} & \frac{1}{2} & 0
\end{pmatrix}, \label{eq:A6modes_alpha}
\end{equation}
where the weights of the weak links are given by
\begin{equation}
    A_{13}^{\infty} = \lim_{r\rightarrow\infty} A_{13} =  e^{-6\alpha}/2. 
\end{equation}

In the limits $\alpha\rightarrow\infty$ and $r\rightarrow\infty$, the adjacency matrix reduces to
\begin{equation}
    \mathbf{A_0} = \lim_{\alpha, r \to \infty} \mathbf{S_C}\mathbf{S_A}^{-1} 
              = \lim_{\alpha, r \to \infty} \mathbf{S_D}\mathbf{S_B}^{-1}. 
\end{equation}
For a six-mode system, $\mathbf{A}_0$ is given by Eq.~(\ref{eq:A6modes}). This adjacency matrix has the same connectivity as $\mathbf{A}(r)$ derived in Sec.~\ref{sec:protocol}. In contrast, $\mathbf{A}_0$ is independent of the squeezing parameter $r$.

\section{Average Nullifier for System Sizes $N=3$--$6$ \label{appendix:e}}
\subsection*{Average Nullifier Based on $\mathbf{A}(r)$}
Using the adjacency matrix $\mathbf{A}(r)=\tanh(2r)\mathbf{A_0}$, the average nullifier can be derived for arbitrary system size $N$ following the procedure outlined in Sec.~\ref{sec:protocol}. The result for $N=3$ is given in Eq.~(\ref{eq:Nu_ave}). For $N=4$, we obtain
\begin{equation}
    S_{\rm nu} = \frac{e^{-2\alpha}}{8} \left( 1 + 3 \sech^2{(2r)} \right).
\end{equation}
For $N=5$, we obtain 
\begin{equation}
    S_{\rm nu} = \frac{e^{-2\alpha}}{10} \left( 1 + 4 \sech^2{(2r)} \right).
\end{equation}
For $N=6$, we obtain 
\begin{equation}
    S_{\rm nu} = \frac{e^{-2\alpha}}{12} \left( 1 + 5 \sech^2{(2r)} \right).
\end{equation}

\subsection*{Average Nullifier Based on $\mathbf{A}(r, \alpha)$}
Using the adjacency matrix $\mathbf{A}(r,\alpha)$ introduced in Appendix~\ref{appendix:d}, we derive the average nullifier following the procedure outlined in Sec.~\ref{sec:protocol}. For $N=3$ (corresponding to six mechanical modes), we obtain
\begin{eqnarray}
    S_{\rm nu}^{r,\alpha} &=& \frac{1}{3} \left[ e^{-2\alpha} \left(1+ 2\sech^{2}(2r)\right) \right. \nonumber\\ 
    && + e^{-6\alpha} (1+ \sech^{2}(2r))\tanh^2(2r) \nonumber\\
    && \left. + e^{-10\alpha} \tanh^4(2r)\right]. 
\end{eqnarray}
For $N=4$, we obtain
\begin{eqnarray}
    S_{\rm nu}^{r,\alpha} &=& \frac{1}{4} \left[ e^{-2\alpha} \left(1+ 3\sech^{2}(2r)\right) \right. \nonumber\\ 
    && + e^{-6\alpha} (1+ 2\sech^{2}(2r))\tanh^2(2r) \nonumber\\
    && + e^{-10\alpha} (1+ \sech^{2}(2r))\tanh^4(2r) \nonumber\\
    && \left. + e^{-14\alpha} \tanh^{6}(2r) \right].
\end{eqnarray}
For $N=5$, we obtain
\begin{eqnarray}
    S_{\rm nu}^{r,\alpha} &=& \frac{1}{5} \left[ e^{-2\alpha} \left(1+ 4\sech^{2}(2r)\right) \right. \nonumber\\ 
    && + e^{-6\alpha} (1+ 3\sech^{2}(2r))\tanh^2(2r) \nonumber\\
    && + e^{-10\alpha} (1+ 2\sech^{2}(2r))\tanh^4(2r) \nonumber\\
    && + e^{-14\alpha} (1+ \sech^{2}(2r))\tanh^6(2r) \nonumber\\
    && \left. + e^{-18\alpha} \tanh^{8}(2r) \right].
\end{eqnarray}
For $N=6$, we obtain
\begin{eqnarray}
    S_{\rm nu}^{r,\alpha} &=& \frac{1}{6} \left[ e^{-2\alpha} \left(1+ 5\sech^{2}(2r)\right) \right. \nonumber\\ 
    && + e^{-6\alpha} (1+ 4\sech^{2}(2r))\tanh^2(2r) \nonumber\\
    && + e^{-10\alpha} (1+ 3\sech^{2}(2r))\tanh^4(2r) \nonumber\\
    && + e^{-14\alpha} (1+ 2\sech^{2}(2r))\tanh^6(2r) \nonumber\\
    && + e^{-18\alpha} (1+ \sech^{2}(2r))\tanh^8(2r) \nonumber\\
    && \left. + e^{-22\alpha} \tanh^{10}(2r) \right]. 
\end{eqnarray}
Compared with the average nullifier $S_{\rm nu}$ derived using the adjacency matrix $\mathbf{A}(r)$ in Sec.~\ref{sec:protocol}, the above expressions contain additional terms involving higher powers of the exponential factor, such as $e^{-6\alpha}$. We also find that, for all of the above results ($N=3$--$6$), the ratio $S_{\rm nu}^{r,\alpha}/S_{\rm nu}\geq 2$ for all values of $r$ and $\alpha$, with equality occurring only at $r=0$.

\subsection*{Average Nullifier Based on $\mathbf{A}_0$}
Using the adjacency matrix $\mathbf{A}_0$, we calculate the average nullifier following the procedure outlined in Sec.~\ref{sec:protocol}. For $N=3$, we obtain
\begin{equation}
    S_{\rm nu,0} = \frac{e^{-2\alpha}}{6} \left[ 
   2 e^{-4r} + 2 e^{-4r} e^{4\alpha} + 1 \right]. \label{eq:Nu_ave_A0}
\end{equation}
For $N=4$, we obtain
\begin{equation}
    S_{\rm nu,0} = \frac{e^{-2\alpha}}{8} \left[ 
   3 e^{-4r} + 3 e^{-4r} e^{4\alpha} + 1 \right].
\end{equation}
For $N=5$, we obtain 
\begin{equation}
    S_{\rm nu,0} = \frac{e^{-2\alpha}}{10} \left[ 
   4 e^{-4r} + 4 e^{-4r} e^{4\alpha} + 1 \right].
\end{equation}
For $N=6$, we obtain 
\begin{equation}
    S_{\rm nu,0} = \frac{e^{-2\alpha}}{12} \left[ 
   5 e^{-4r} + 5 e^{-4r} e^{4\alpha} + 1 \right].
\end{equation}

\begin{figure}
    \centering
    \includegraphics[width=8.5cm, clip]{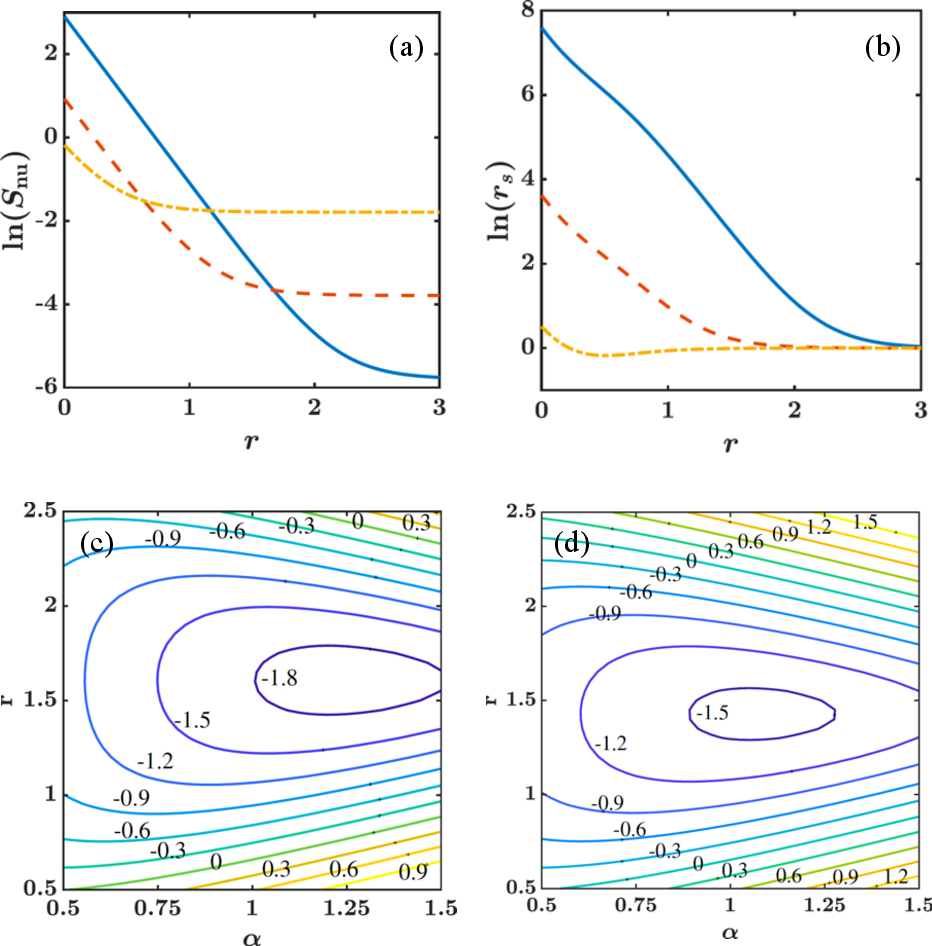}
    \caption{(a) Logarithm of the average nullifier $\ln(S_{\rm nu,0})$ vs $r$.
    (b) Logarithm of the ratio $r_s$ vs $r$. In both panels, $\alpha=2$, $1$, and $0$ correspond to the solid, dashed, and dash-dotted curves, respectively. Other parameters are the same as those in Fig.~\ref{fig3}.
    (c) and (d) Contour plots of $\ln(S_{\rm nu,0})$ vs $\alpha$ and $r$ for $\kappa=0.02$ and $0.04$, respectively. Here $n_{\rm th}=2$ and $\gamma=5\times10^{-4}$.}
    \label{fig7}
\end{figure}

For illustration, Fig.~\ref{fig7}(a) plots the logarithm of the average nullifier in Eq.~(\ref{eq:Nu_ave_A0}) for $N=3$ as a function of the two-mode squeezing parameter $r$. Although $S_{\rm nu,0}$ also decreases monotonically with increasing $r$, it is significantly larger than $S_{\rm nu}$ in Eq.~(\ref{eq:Nu_ave}) at small values of $r$ for nonzero $\alpha$. We define the ratio between the two nullifiers as $r_s=S_{\rm nu,0}/S_{\rm nu}$. Figure~\ref{fig7}(b) shows the logarithm of $r_s$ as a function of $r$. For nonzero initial momentum squeezing $\alpha$, one finds that $r_s\gg1$ for small and moderate values of $r$. Only in the case of $\alpha=0$, corresponding to the absence of initial momentum squeezing, can the ratio $r_s$ become smaller than one over a small range of $r$.
Figures~\ref{fig7}(c) and \ref{fig7}(d) present contour plots of $\ln(S_{\rm nu,0})$ in the presence of dissipation and thermal noise, using the same parameters as in Fig.~\ref{fig4}(e) and (f). The dependence of $S_{\rm nu,0}$ on the squeezing parameters is qualitatively similar to that of $S_{\rm nu}$, although the corresponding nullifier values are generally larger.

\section{Matrix $\mathbf{C}$ for Master Equation \label{appendix:f}}
Following the approach of Ref.~\cite{KogaPRA2012}, we define a vector $\mathbf{c}_m$ for each damping channel described by the Liouvillian superoperator $\mathcal{L}_m$ in the master equation. For a generic channel, $\mathcal{L}_m(\rho)=L_m\rho L_m^\dagger-\frac{1}{2}L_m^\dagger L_m\rho-\frac{1}{2}\rho L_m^\dagger L_m$, with $L_m=\mathbf{c_{m}}^T\mathbf{\widetilde{R}}$. For a cavity mode $a_n$ ($n\in[1,N]$), the Liouvillian is 
\begin{equation}
\mathcal{L}_{an}[\rho] = \frac{\kappa}{2} \left( 2 a_n \rho a_n^\dagger - \rho a_n^\dagger a_n - a_n^\dagger a_n \rho \right),
\end{equation}
where $\kappa$ is the cavity damping rate. Using $a_n = \frac{1}{\sqrt{2}}(Q_{an} + i P_{an})$, the corresponding $L_m$ operator is $L_m = \sqrt{\frac{\kappa}{2}}(Q_{an} + i P_{an})$, from which the vector $\mathbf{c}_m$ can be derived directly.
For a mechanical mode $b_n$ ($n\in[1,2N]$), the Liouvillian is
\begin{align}
\mathcal{L}_n[\rho] &= \frac{\gamma (n_{\rm th}+1)}{2} \left( 2 b_n \rho b_n^\dagger - \rho b_n^\dagger b_n - b_n^\dagger b_n \rho \right) \nonumber\\
&\quad + \frac{\gamma n_{\rm th}}{2} \left( 2 b_n^\dagger \rho b_n - \rho b_n b_n^\dagger - b_n b_n^\dagger \rho \right),
\end{align}
where $\gamma$ is the mechanical damping rate and $n_{\rm th}$ is the thermal phonon number. This Liouvillian can be separated into two channels: the first channel corresponds to the $(n_{\rm th}+1)$ term with $L_m = \sqrt{\frac{\gamma (n_{\rm th}+1)}{2}}(Q_n + i P_n)$, and the second channel corresponds to the $n_{\rm th}$ term with $L_m = \sqrt{\frac{\gamma n_{\rm th}}{2}}(Q_n - i P_n)$. The vectors $\mathbf{c_{m}}$ associated with these channels can then be derived accordingly.

The matrix $\mathbf{C}$ is defined as $\mathbf{C} = (\mathbf{c_1}, \mathbf{c_2}, \cdots, \mathbf{c_M})^T$, where $M$ is the total number of damping channels. For a phononic quantum network with $2N$ mechanical modes and $N$ cavity modes, there are a total of $M = 5N$ damping channels: $4N$ channels from the mechanical modes (two channels for each mode) and $N$ channels from the cavity modes. Each vector $\mathbf{c_m}$ has $6N$ elements corresponding to all mechanical and cavity quadratures. Consequently, $\mathbf{C}$ is a $5N \times 6N$ matrix. 

\section{Covariance Matrix after Measurements \label{appendix:g}}
We follow the approach of Ref.~\cite{WeedbrookRMP2012GaussianQI} to derive the covariance matrix of the remaining modes after performing a quadrature measurement (displacement or momentum) on a single mode. Given the initial covariance matrix of the quantum network $\mathbf{V}$, we partition it according to the measured mode and the remaining modes as
\begin{equation}
\mathbf{V} =
\begin{pmatrix}
\mathbf{V}_{\rm Re} & \mathbf{V}_{\rm col} \\
\mathbf{V}_{\rm col}^{T} & \mathbf{V}_{\rm M}
\end{pmatrix},
\end{equation}
where $\mathbf{V}_{\rm M}$ is the covariance matrix of the measured mode, $\mathbf{V}_{\rm Re}$ is the covariance matrix of the remaining modes, and $\mathbf{V}_{\rm col}$ contains the correlations between these two subsystems.
After the measurement, the covariance matrix of the remaining modes becomes
\begin{equation}
\mathbf{V}_{\rm Re} \rightarrow \mathbf{V}_{\rm Re} - \mathbf{V}_{\rm col} (\Pi \mathbf{V}_{\rm M} \Pi)^{-1} \mathbf{V}_{\rm col}^{T}, \label{eq:Vmeasure}
\end{equation}
where the diagonal matrix $\Pi = {\rm Diag}(1,0)$ for a displacement measurement, $\Pi = {\rm Diag}(0,1)$ for a momentum measurement, and $(\Pi \mathbf{V}_{\rm M} \Pi)^{-1}$ denotes the pseudoinverse operation. 
By applying the transformation in Eq.~(\ref{eq:Vmeasure}) sequentially for each measurement described in Sec.~\ref{sec:entanglementtransfer}, we obtain the final covariance matrix of the two remaining modes $b_{N+1}$ and $b_{2N}$.


\end{document}